\documentstyle[12pt,psfig,draft,lscape,array,amssymb]{nature-ejb}

\def\simlt{\mathrel{\hbox{\rlap{\hbox{\lower4pt\hbox{$\sim$}}}\hbox{$<$}}}}
\def\simgt{\mathrel{\hbox{\rlap{\hbox{\lower4pt\hbox{$\sim$}}}\hbox{$>$}}}}

\def\ale{\mathrel{\hbox{\rlap{\hbox{\lower4pt\hbox{$\sim$}}}\hbox{$<$}}}}
\def\age{\mathrel{\hbox{\rlap{\hbox{\lower4pt\hbox{$\sim$}}}\hbox{$>$}}}}

\def\nodata{---}

\def\ra#1#2#3{#1$^{\rm h}$#2$^{\rm m}$#3$^{\rm s}$}
\def\dec#1#2#3{$#1^\circ#2'#3''$}

\newcommand{\sgr}{SGR\,1806$-$20}

\begin{document}

\title{\bf Discovery of a Radio Source following the 27 \\ December 2004 Giant Flare from SGR\,1806$-$20}

\author{ 
P.~B.~Cameron\affiliation[1]{\scriptsize Division of Physics, Mathematics and Astronomy,
  105-24, California Institute of Technology, Pasadena, CA 91125, USA},
P.~Chandra\affiliation[2]{\scriptsize Tata Institute of Fundamental Research, Mumbai 400 005,
  India}$^,$\affiliation[3]{\scriptsize Joint Astronomy Programme, 
  Indian Institute of Science, Bangalore 560 012, India},
A.~Ray\affiliationmark[2],
S.~R.~Kulkarni\affiliationmark[1],
D.~A.~Frail\affiliation[4]{\scriptsize National Radio Astronomy Observatory, Socorro, NM 87801, USA},
M.~H.~Wieringa\affiliation[5]{\scriptsize Australia Telescope National Facility, CSIRO,
P.O. Box 76, Epping NSW 1710, Australia},
E.~Nakar\affiliation[6]{\scriptsize Theoretical Astrophysics 130-33, California Institute of
Technology, Pasadena, CA\,91125, USA},
E.~S.~Phinney\affiliationmark[6],
Atsushi Miyazaki\affiliation[7]{\scriptsize Shanghai Astronomical Observatory 80 Nandan Road 
Shanghai 200030, China},
Masato Tsuboi\affiliation[8]{\scriptsize Nobeyama Radio Observatory,National Astronomical 
Observatory of Japan Minamisaku, Nagano 384-1305, Japan},
Sachiko Okumura\affiliationmark[8],
N.~Kawai\affiliation[9]{\scriptsize Department of Physics Tokyo Institute of Technology 
Ookayama 2-12-1, Meguro-ku,Tokyo 152-8551, Japan},
K.~M.~Menten\affiliation[10]{\scriptsize Max Planck Institut f\"ur Radioastronomie, Auf dem H\"ugel
69; 53121 Bonn, Germany} \&\
F.~Bertoldi\affiliation[11]{\scriptsize University of Bonn, Auf dem H\"uegel 71, 53121 Bonn, Germany}
}
\date{\today}{}
\headertitle{\sgr\ Fading Radio Source}
\mainauthor{Cameron et al.}

\summary{
Over a decade ago it was established that the remarkable high energy
transients, known as soft gamma-ray repeaters (SGRs), are a Galactic
population\cite{kf93,mtk+94} and originate from neutron stars with
intense ($\simlt 10^{15}\,$G) magnetic fields (``magnetars''\cite{wt04}).
On 27 December 2004 a giant flare\cite{GCN2920} (fluence\cite{GCN2936}
$\age 0.3\,$erg cm$^{-2}$) was detected from \sgr.  Here we report
the discovery of a fading radio counterpart.  We began a monitoring
program from 0.2\,GHz to 250\,GHz and obtained a high resolution
21-cm radio spectrum which traces the intervening interstellar
neutral Hydrogen clouds.  Analysis of the spectrum yields the first
direct distance measurement of \sgr. The source is located at a 
distance greater than 6.4\,kpc and we argue that it is nearer than 9.8\,kpc.  
If true, our distance estimate lowers the total energy of the explosion 
and relaxes the demands on theoretical models. 
The energetics and the rapid decay of the
radio source are not compatible with the afterglow model that is
usually invoked for gamma-ray bursts. 
Instead we suggest that the rapidly decaying radio emission arises from the debris
ejected during the explosion.
}

\maketitle


On 3 January 2005 we observed \sgr\
with the Very Large Array (VLA) and identified and promptly
reported\cite{GCN2928} a new radio source at right ascension
$\alpha_{J2000}=$\ra{18}{08}{39.34} and declination
$\delta_{J2000}=$\dec{-20}{24}{39.7} (with an uncertainty of $\pm
0.1$'' in each coordinate) coincident with the quiescent X-ray
counterpart\cite{kfk+02}. In Table~\ref{tab:fluxes} we report the
results of a subsequent monitoring program undertaken with the VLA, the
Giant Metrewave Radio Telescope (GMRT), the Australia Telescope
Compact Array (ATCA), the Nobeyama Millimeter Array (NMA) and the
Institut de Radioastronomie Millim$\acute{\rm e}$trique (IRAM) 30m Telescope. 

\begin{figure}
\centerline{\psfig{file=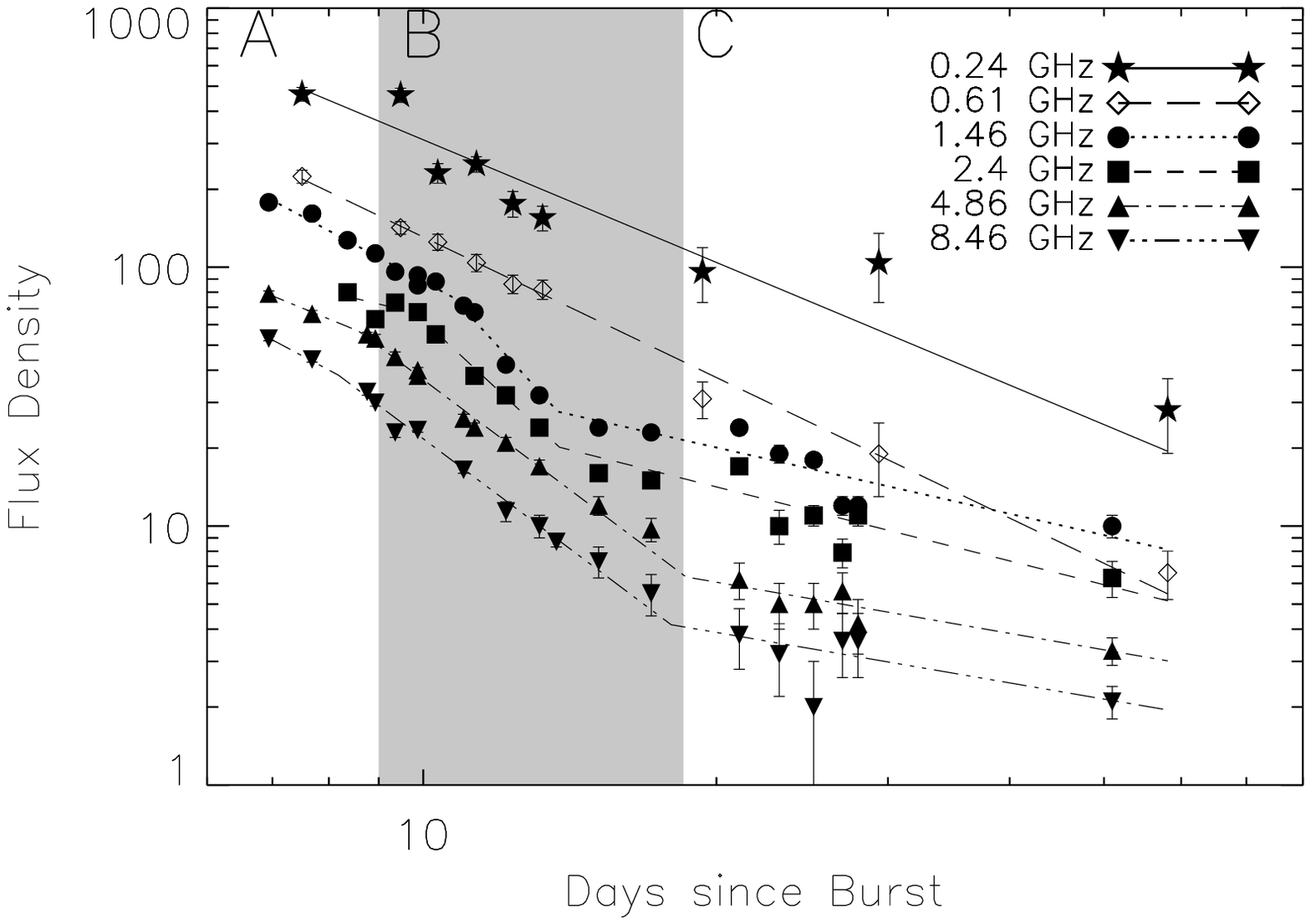,width=6in,angle=0}}
\bigskip
\bigskip
\caption[]{}
\label{fig:LightCurves}
\end{figure}

The radio source decays in all frequency bands, but the behaviour
is complex (Figure~\ref{fig:LightCurves}).  At each band we model
the flux by a power law, $S_\nu(t)\propto t^\alpha$, but
allow for changes in the temporal indices $\alpha$ (``breaks'') at two epochs. These
breaks are clearly seen in our highest signal-to-noise ratio data.
Following the first break (9 days, postburst) the light curve steepens 
to $-4 \simlt \alpha \simlt -3$.
The radio source\cite{fkb99} from SGR\,1900+14 following the 27 August 1998 giant flare\cite{fhd+01}
showed a similar rapid decay at 8\,GHz.
Subsequently around day 14 the light curve flattens to
$\alpha\sim -1$.  At any given epoch, the radio spectrum can be
modeled by a power law, $S_\nu \propto \nu^{\beta}$.  The spectral
index, $\beta$, steepens with time, changing from $\sim -0.7$ to $-0.9$ (see
Figure~\ref{fig:LightCurves} and Supplemental Information).  

We confirm claims that the source is resolved\cite{GCN2933} by an independent
analysis.  We find that it is elongated with a major axis $\theta\sim
77$\,milliarcsecond (mas) and an axial ratio of 2:1 (Table~\ref{tab:size}).
We considered four expansion models, $\theta\propto t^{s}$ with 
unconstrained $s$ and  plausible expansion models ($s=0$, 2/5, and 1). 
The best fit model corresponds to no expansion ($s=0.04\pm 0.15$). However,
due to the limited range of our observations we prefer not to model the dynamics of the
explosion.

\begin{figure}
\centerline{\psfig{file=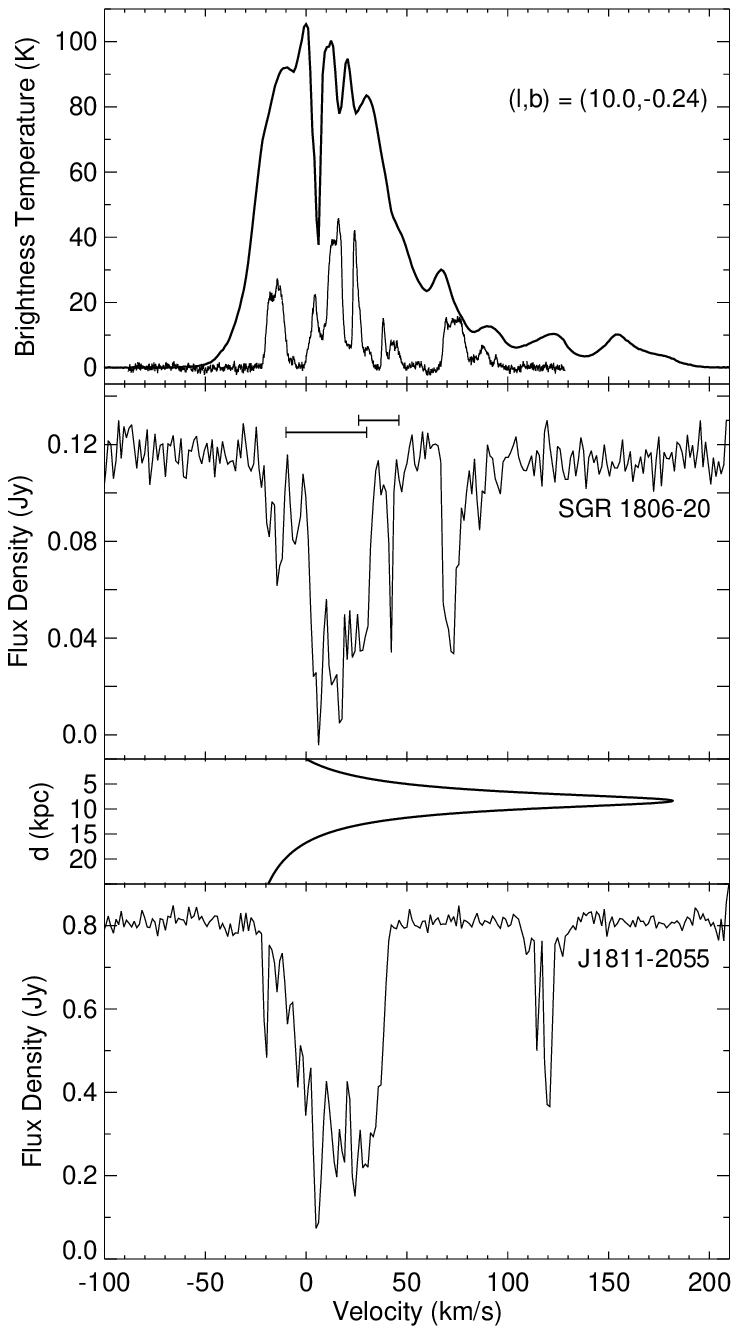,width=4.5in,angle=0}}
\bigskip
\bigskip
\caption[]{}
\label{fig:hone}
\end{figure}

We took advantage of the brightness of the radio source and obtained
a high resolution spectrum (Figure~\ref{fig:hone}, second panel) centered around
the 21-cm line of atomic Hydrogen (HI).   Intervening interstellar
clouds appear as absorption features in the spectrum. These clouds
are expected to participate in the rotation of the Galaxy and the
absorption features allow us to infer ``kinematic'' distance estimates.
Such estimates have several complications. First, in the inner Galaxy the
radial velocity curve is double-valued (see Figure~\ref{fig:hone}, third panel)
leading to a ``near'' distance estimate
(d$_l$) and a ``far'' distance estimate (d$_u$) for each velocity. Second, in
some directions, there are features with non-circular motion e.g.
the ``3-kpc expanding arm'' and the ``$-30$ km\,s$^{-1}$ spiral
arm''\cite{ce04}.  Finally, in the innermost part of the Galaxy
there is a deficit of cold gas\cite{kjbd02}.

Significant HI absorption toward \sgr\ is seen over the velocity
range $-$20 km\,s$^{-1}$ to +85 km\,s$^{-1}$ (Figure~\ref{fig:hone}, second panel).  
There is also a weak (2.5-$\sigma$) absorption feature coincident in velocity with a
clearly detected $^{12}$CO(1-0) emission feature identified\cite{cwd+97} as MC94 (Figure~\ref{fig:hone}, first panel).
Adopting a simple Galactic rotation curve with a circular velocity
$\Theta_\circ=220$ km\,s$^{-1}$ and a Galactic center distance
R$_\circ$=8.5\,kpc, the near distance to \sgr\ (for V$_{LSR}$=95 
km\,s$^{-1}$) is d$_l$=6.4\,kpc.

The two HI emission clouds seen at velocities above 100 km\,s$^{-1}$
toward \sgr\ (Figure~\ref{fig:hone}, first panel), with no corresponding HI absorption, may be used to
infer an upper limit to the distance provided that we can be
reasonably certain that cold neutral gas exists at these velocities.
The HI absorption spectrum toward the nearby ($\Delta\theta=0.77^\circ$)
extragalactic source J1811$-$2055 shows a strong and broad absorption
feature between 110 and 130 km\,s$^{-1}$ (Figure~\ref{fig:hone}, fourth panel).
The only HI emission in this direction\cite{hb97} above 60 
km\,s$^{-1}$ corresponds to an HI cloud at this same velocity. This
feature can be traced in absorption toward several other
extragalactic radio sources in this direction \cite{gd89}, suggesting
that cold gas at $\sim$120 km\,s$^{-1}$ is widespread.  
Adopting the same Galactic rotation curve as above, the
absorbing cloud at +120 km\,s$^{-1}$ can either be located at 7\,kpc
or 9.8\,kpc (see Figure~\ref{fig:hone}, third panel).
We thus suggest an upper limit to the distance, d$_u$=9.8\,kpc.  

Our new distance estimate is smaller than previous (indirect) values\cite{fnk04,ce04} 
of 12 to 15\,kpc. Accepting our estimate has several important implications.
It results in a reduction of the total energy released ($\propto d^2$) as
well as the rate of such events in nearby galaxies\cite{ngp+05} ($\propto d^3$), 
and calls into question the association\cite{fmc+99} of \sgr\ 
with a star cluster along the same line-of-sight.
Therefore, claims that magnetars originate from more massive stars than normal
neutron stars\cite{gmo+05} may be called into question.

Next we consider the energetics of the material giving rise to 
the radio emission.
As with many other radio sources, the power law spectrum can be
attributed to energetic electrons with a power law energy distribution
($dN/d\gamma \propto \gamma^{-p}$; here $\gamma$ is the Lorentz
factor of electrons and we measure $p = 2.24 \pm 0.04$ on day 7, 
which as is a typical value for strong shocks) 
which gyrate in a magnetic field and emit synchrotron radiation.  
We apply the minimum energy formulation for synchrotron sources\cite{Pac70,sr77}
to the radio spectrum (from 0.2\,GHz to 100\,GHz) of 3 January 2005 and find the
energy of the radio emitting source and the associated magnetic
field strength are $U_{min} \sim  10^{43} d_{10}^{17/7}
\theta_{75}^{9/7}$ erg and $B_{min} \sim 13 d_{10}^{-2/7}
\theta_{75}^{-6/7}$ mG; here, the distance is 10$d_{10}$\,kpc and the
angular diameter is 75$\theta_{75}$\,mas (Table~\ref{tab:size}).

Evidently the amount of energy released in the $\gamma$-ray flare,
$E_{\gamma,{\rm iso}}\simgt 4\times 10^{45}\,$erg\,s$^{-1}$ (assuming
unbeamed, isotropic emission),
substantially exceeds $U_{\rm min}$. 
In contrast, the ratio $U_{\rm min}/E_{\gamma,{\rm iso}}$ is unity for 
GRBs and as a result the lower energy and
longer duration emission is correctly regarded as arising from the
shock of the circumburst medium (the ``afterglow'' model).  Thus,
based solely on energetics, there is no {\em prima facie} reason to
suggest that the radio source is the afterglow of the $\gamma$-ray
flare. 

Furthermore, as discussed above, the radio emission decays quite
rapidly 9 days after the burst.
Such a rapid decay is incompatible
with the afterglow model (in the non-relativistic limit)  for which
we expect\cite{fwk00} $\alpha=3\beta+0.6$.  We conclude (in contrast
to refs.~\pcite{cw03}, \pcite{nps05} and \pcite{wwf+05}) that  the
radio emission must be powered by something very different from
that which produced the $\gamma$-ray emission. 

In summary, the radio emission can be described by two components:
a rapidly decaying component and a slowly decaying component.
The latter
becomes detectable when the former has decayed significantly.  The
rapid decay is phenomenologically similar to that seen from accreting
Galactic sources (e.g. ref.  \pcite{hrh+00}) -- the so-called
``plasmon'' model framework in which the radio emission arises from
a ball of electrons and magnetic field which are initially shocked and
then cool down by expansion. 
We make the specific suggestion that
the radio emission up until about 2 weeks is a result of the shocking
of the debris given off in the explosion (the ``reverse shock'').
In this framework the slowly decaying component is the emission
arising from the forward shock as the ejecta slams into the circumburst
medium. A requirement of this suggestion is that the energy inferred
in the slowly decaying component should be comparable to $U_{\rm min}$.
Separately, we note that the comparable ratios $U_{min}/E_{\gamma,{\rm iso}}$
and the temporal and spectral similarities of 
the giant flares from \sgr\ and SGR\,1900$+$14 suggest a common mechanism for launching
these flares and similar circumstellar environments.

Regardless of the suggestions and speculations, it is clear that radio afterglow is telling us something
entirely different from that revealed by the $\gamma$-ray emission.
If our suggestion of a reverse shock origin is correct then radio
observations allow us to probe the ejecta. Taken together
it appears that rapid and intense radio monitoring of
such flares will be highly fruitful in the future.

\bibliographystyle{nature-pap}
\bibliography{journals,magnetar}

\bigskip
\noindent
{\bf Supplementary Information} accompanies the paper on
{\bf www.nature.com/nature.}

\begin{acknowledge}
ATCA is funded by the Commonwealth of Australia for operation as a
National Facility managed by CSIRO.  We thank K. Newton-McGee and B.
Gaensler for scheduling and performing observations with the ATCA.
GMRT is run by the National Centre for Radio Astrophysics-Tata
Institute of Fundamental Research, India.  We thank the GMRT staff and
in particular C. H. Ishwara-Chandra and D. V. Lal for help with
observations and analysis.  The VLA is a facility of the National
Science Foundation operated under cooperative agreement by Associated
Universities, Inc.  NMA is a branch of the National Astronomical
Observatory, National Institutes of Natural Sciences, Japan.  IRAM is
supported by INSU/CNRS (France), MPG (Germany) and IGN (Spain).  We
thank A.  Weiss from IRAM for help with the observations.  We
gratefully acknowledge S. Corbel, S. S.  Eikenberry and R. Sari for
useful discussions.  Our work is supported in part by the NSF and
NASA.

\bigskip
\noindent
{\bf Correspondence} and requests for materials should be addressed to P.B.C.
(pbc@astro.caltech.edu).

\end{acknowledge}

\clearpage
\vspace{0.5cm}
\centerline{\bf FIGURES \& CAPTIONS}
\vspace{0.5cm}

\noindent
{\bf Table~\ref{tab:fluxes}}: {\small 
Flux density measurements of the transient radio counterpart to
\sgr\ from the VLA, GMRT, NMA, and ATCA as a function of frequency
and time.  The reported errors are 1-$\sigma$.  In addition to these
measurements, we obtained IRAM-30m observations on 8 and 9 January
2005 using MAMBO-2 at 250\,GHz which show no detection with a value
of $0.57 \pm 0.46$~mJy at the position of the radio source.  Finally,
we detect linearly polarized emission from the source at the 1.5\%
to 2.5\% level.  See the Supplemental Information for observational
details.\\ $^a$ The epoch of the flare, $t_0$, was 27.90 December
2004.\\ $^b$ ATCA observations in this column have a frequency of
8.6\,GHz.\\ $^c$ These values represent 2$\sigma$ upper limits.\\
$^d$ The frequency is 1.06\,GHz for the 16.37 January 2005 and 4.01
February 2005 GMRT observations.\\ 
}

\noindent
{\bf Figure~\ref{fig:LightCurves}}: {\small Broadband temporal
behavior of the transient radio source coincident with \sgr.  The
abscissa indicates days elapsed since the giant flare on 27.90
December 2004.  The displayed flux density measurements (denoted
with symbols) were obtained in six frequency bands with the VLA,
GMRT, and ATCA (Table~\ref{tab:fluxes}).  The error bars denote
1-$\sigma$ uncertainties.  With the exception of the 6.1\,GHz data
(which is insufficiently sampled at early and late times and is not
shown), the light curves with $\nu > 1\,$GHz are best fit by power-law
models (shown as lines, $S_\nu \propto t^{\alpha_i}$) with two
breaks at $t_1 \sim 9$ days and $t_2 \sim 15$ days (see
the Supplemental Information for exact values).  The
temporal index varies chromatically in the time before and after
the first break (denoted by regions A and B respectively).  
The exponent value ranges from $-2 \simlt \alpha_A \simlt -1$ and $-4
\simlt \alpha_B \simlt -3$; here the subscript identifies the region
of interest.  After day $\sim 15$ (region C) the source decay
flattens to $\alpha_C \sim -0.9$ at these frequencies which persists
until day 51.
Region B, the period of steep light curve decline, is shaded gray. 
The light curves with $\nu < 1\,$GHz do not show these temporal breaks or late time flattening
Apparently a single power law decay model 
with $\alpha=-1.57 \pm 0.11$ (0.24\,GHz)
and $\alpha=-1.80 \pm 0.08$ (0.61\,GHz)
provides a good statistical description of the data.

Our substantial frequency coverage (over three decades) allows
an excellent characterization of the spectrum.  The spectrum
is consistent with a single power law slope ($S_\nu \propto \nu^\beta$)
at all epochs.  On day 7, before the first temporal break, we find $\beta =
-0.62 \pm 0.02$.  The spectrum steepens to a value of $\beta = -0.76
\pm 0.05$ (day 15), reaching $\beta = -0.9 \pm 0.1$ (days 21--51).\\
}

\noindent
{\bf Table~\ref{tab:size}}: {\small Source size measurements and
   95\% confidence limits of the radio source as measured with the
   VLA at 8.46\,GHz.  The source is clearly resolved at all VLA
   epochs. The best constraints on the source size come from the
   observations which occurred closest to the transit of the
   source on Jan 3rd and Jan 6th.  The result is a source with size
   $\theta \sim 77$ mas with an axial ratio of $\sim 0.5$ and a
   position angle (PA) of 60 degrees (measured clockwise from the
   North).  The flux centroid did not change position within the
   limits of our astrometric accuracy ($\pm$ 100 mas). 

   The best fit model is consistent with no expansion, 
   $s = 0.04 \pm 0.15$ ($\theta(t) \propto t^s$) with a $\chi^2 = 5.3$ 
   with 4 degrees of freedom.  
   Sedov-Taylor ($s=2/5$) and free expansion ($s=1$) model were also fit, and
   yield $\chi^2 = 20.6$ and $\chi^2 = 91$, respectively. These fits
   have  five 
   degrees of freedom.
   See the Supplemental Information for the details of the source size
   measurements.\\
}

\noindent
{\bf Figure~\ref{fig:hone}}: {\small Cold atomic and molecular hydrogen spectra 	
   toward \sgr. These spectra were used to derive a distance estimate for \sgr.
   {\em (Top Panel)}. HI emission
   (upper curve, thick line) in the direction of \sgr, determined
   by averaging two adjacent spectra taken by Hartmann \&
   Burton\cite{hb97} at $l,b$=(10.0$^\circ$,0.0$^\circ$) and
   $l,b$=(10.0$^\circ$,$-0.5^\circ$). The lower curve (thin line)
   is the $^{12}$CO(1$-$0) spectrum (from ref.~\pcite{ce04}).
   For display purposes the brightness temperature  has been
   scaled up by a factor of 11.4.
   {\em (Second Panel).}
   The HI absorption spectrum taken toward \sgr.
   The two horizontal bars illustrate the radial velocity
   measurements\cite{fnk04,ce04} of the nearby star LBV 1806$-$20
   (36$\pm$10\,km\,s$^{-1}$ and 10 $\pm$20\,km\,s$^{-1}$).  
   The absorption spectra were
   made with the Very Large Array on 4 January 2005, using a 1.56
   MHz bandwidth in both hands of polarization centered 50 
   km\,s$^{-1}$ with respect to the local standard of rest.  The bandwidth
   was divided into 256 channels each 6.1 kHz in width, or a velocity
   resolution of 1.3 km\,s$^{-1}$ covering a velocity range of $-$115
   km\,s$^{-1}$ to $+$215 km\,s$^{-1}$.
   {\em (Third Panel).} The distance as a function of radial velocity adopting a
   simple Galactic rotation curve with a circular velocity
   $\Theta_\circ=220$ km\,s$^{-1}$ and a galactic center distance
   R$_\circ$=8.5\,kpc.  
   {\em (Bottom Panel).} The HI absorption spectrum
   of nearby extragalactic source J1811$-$2055 at
   $l,b$=(9.8$^\circ$,$-1.0^\circ$).  \\

The lower limit to the distance is firmly established by the 95\,km\,s$^{-1}$
absorption feature from MC94 (see Text). 
An upper limit to the distance of the SGR is suggested by the {\it
absence} of strong absorption at +120 km s$^{-1}$, seen toward
J1811$-$2055 and several other extragalactic radio sources in this
direction\cite{gd89}. One could argue that the
120\,km\,s$^{-1}$ cold cloud is small and not present along the
line-of-sight to the \sgr.  However, this hypothesis also requires
the absence of any other cloud between 95\,km\,s$^{-1}$ (distance
of 6.4\,kpc) and 86\,km\,s$^{-1}$ (distance of 10.6\,kpc).
The mean absorption coefficient drops in the inner Galaxy ($R \simlt 4.5$)
kpc, giving a mean free path between clouds of 2.3 kpc\cite{gd89}.
The distance interval from 6.4\,kpc to 10.6\,kpc
corresponds to $\sim 1.8$ mean free paths. So, the probability of finding no
clouds in this gap is 16\%. 
Thus our upper limit of 9.8\,kpc is not a certainty but quite likely.
}

\begin{landscape}
\begin{table}
\begin{center}
\setlength{\extrarowheight}{-0.090in}
\begin{tabular}{>{\scriptsize}l >{\scriptsize}l >{\scriptsize}c >{\scriptsize}c >{\scriptsize}c >{\scriptsize}c >{\scriptsize}c >{\scriptsize}c >{\scriptsize}c >{\scriptsize}c >{\scriptsize}c}
\hline
\hline
& & $\Delta t^a$&S$_{0.24}$&S$_{0.610}$&S$_{1.46}$&S$_{2.4}$&S$_{4.86}$&S$_{6.1}$&S$_{8.46}^b$&S$_{102}$\\
Epoch & Telescope & (days)&  (mJy)& (mJy)& (mJy)& (mJy)& (mJy)& (mJy)& (mJy)&(mJy)\\
\hline
3.84 Jan 2005 & VLA & 6.94 & \nodata       & \nodata      & 178 $\pm$ 4 & \nodata    & 79 $\pm$ 2    & \nodata   & 53 $\pm$ 1    & \nodata\\
4.17 Jan 2005 & NMA & 7.27 & \nodata       & \nodata      & \nodata     & \nodata    & \nodata       & \nodata   & \nodata       & 16.3 $\pm$ 5.6\\
4.41 Jan 2005 & GMRT& 7.51 & 466 $\pm$ 28  & 224 $\pm$ 13 & \nodata     & \nodata    & \nodata       & \nodata   & \nodata       & \nodata  \\
4.59 Jan 2005 & VLA & 7.69 & \nodata       & \nodata      & 161 $\pm$ 4 & \nodata & 66 $\pm$ 2    & \nodata   & 44 $\pm$ 1    & \nodata \\
5.26 Jan 2005 & ATCA& 8.36 & \nodata       & \nodata      & 127 $\pm$ 3 & 80 $\pm$ 2 & \nodata       &  \nodata  & \nodata       & \nodata\\
5.66 Jan 2005 & VLA & 8.76 & \nodata       & \nodata      & \nodata     & \nodata    & 55 $\pm$ 1    & \nodata   & 33 $\pm$ 1    & \nodata \\
5.85 Jan 2005 & ATCA& 8.93 & \nodata       & \nodata      & 113 $\pm$ 3 & 63 $\pm$ 2 & 53  $\pm$ 2   & \nodata   & 30 $\pm$ 1    & \nodata \\
6.26 Jan 2005 & ATCA& 9.36 & \nodata       & \nodata      & 96 $\pm$ 3  & 73 $\pm$ 2 & 45 $\pm$ 2    &  \nodata  & 23 $\pm$ 1    & \nodata \\ 
6.38 Jan 2005 & GMRT& 9.48 & 462 $\pm$ 29  & 142 $\pm$ 8  & \nodata     & \nodata    & \nodata       & \nodata   & \nodata       & \nodata \\
6.77 Jan 2005 & VLA & 9.87 & \nodata       & \nodata      & 93 $\pm$ 2  & \nodata    & 38 $\pm$ 1    & \nodata   & 23.5 $\pm$ 0.5& \nodata \\
6.77 Jan 2005 & ATCA& 9.87 & \nodata       & \nodata      & 85 $\pm$ 3  & 67 $\pm$ 2 &  40 $\pm$ 1   & 32 $\pm$ 1& \nodata       & \nodata \\
7.20 Jan 2005 & ATCA& 10.30& \nodata       & \nodata      & 88 $\pm$ 2  & 55 $\pm$ 1 & \nodata       &  \nodata  & \nodata       & \nodata \\ 
7.25 Jan 2005 & GMRT& 10.35& 231 $\pm$ 20  & 125 $\pm$ 9  & \nodata     & \nodata    & \nodata       & \nodata   & \nodata       & \nodata\\
7.90 Jan 2005 & VLA & 11.00& \nodata       & \nodata      & 71 $\pm$ 2  & \nodata    & 26 $\pm$ 1    & \nodata   & 16.5 $\pm$ 0.5& \nodata\\
8.19 Jan 2005 & ATCA& 11.29& \nodata       & \nodata      & 67 $\pm$ 3  & 38 $\pm$ 2 & 24 $\pm$ 1    & 20 $\pm$ 1& \nodata       & \nodata\\
8.24 Jan 2005 & GMRT& 11.34& 250 $\pm$ 17  & 104 $\pm$ 8  & \nodata     & \nodata    & \nodata       & \nodata   & \nodata       & \nodata\\ 
9.06 Jan 2005 & ATCA& 12.16& \nodata       & \nodata      & 42 $\pm$ 2  & 32 $\pm$ 1.5& 21 $\pm$ 1   & \nodata   & 11.4 $\pm$ 1  & \nodata\\ 
9.26 Jan 2005 & GMRT& 12.36& 176 $\pm$ 20  & 86 $\pm$ 7   & \nodata     & \nodata    & \nodata       & \nodata   & \nodata       & \nodata\\
10.07 Jan 2005& ATCA& 13.16& \nodata       & \nodata      & 32 $\pm$ 2  & 24  $\pm$ 1& 17  $\pm$ 1   & \nodata   & 10 $\pm$ 1    & \nodata\\
10.16 Jan 2005& GMRT& 13.26& 155 $\pm$ 17  & 82 $\pm$ 7   &\nodata      & \nodata    & \nodata       & \nodata   & \nodata       & \nodata\\
10.60 Jan 2005& VLA & 13.70& \nodata       & \nodata      & \nodata     & \nodata    & \nodata       & \nodata   & 8.7 $\pm$ 0.4 & \nodata\\
12.00 Jan 2005& NMA & 15.10& \nodata       & \nodata      & \nodata     & \nodata    & \nodata       & \nodata   & \nodata       & 7.16$^c$\\
12.04 Jan 2005& ATCA& 15.14& \nodata       & \nodata      & 24 $\pm$ 1.5& 16 $\pm$ 1 & 12 $\pm$ 1    & 9.3 $\pm$ 1& 7.3 $\pm$ 1  & \nodata\\
13.00 Jan 2005& NMA & 16.10& \nodata       & \nodata      & \nodata     & \nodata    & \nodata       & \nodata   & \nodata       & 5.50$^c$\\
14.04 Jan 2005& ATCA& 17.14& \nodata       & \nodata      & 23 $\pm$ 1  & 15 $\pm$ 1 & 9.7 $\pm$ 1   & 7.3 $\pm$ 1& 5.5 $\pm$ 1  & \nodata\\
16.25 Jan 2005& GMRT& 19.35& 96 $\pm$ 23   & 31 $\pm$ 5   &\nodata      & \nodata    & \nodata       & \nodata    & \nodata        & \nodata\\
16.37 Jan 2005& GMRT& 19.47& \nodata       & \nodata      & 20 $\pm$ 2$^d$& \nodata  & \nodata       & \nodata    & \nodata        & \nodata\\
18.01 Jan 2005& ATCA& 21.11& \nodata       & \nodata     & 24 $\pm$ 1.5 & 17 $\pm$ 1 & 6.2 $\pm$ 1   & 4.7 $\pm$ 1& 3.8$\pm$ 1 & \nodata\\
20.10 Jan 2005& ATCA& 23.20& \nodata       & \nodata     & 19 $\pm$ 1.5 & 10 $\pm$ 1.5& 5  $\pm$ 1   & \nodata    & 3.2$\pm$ 1 & \nodata\\
22.07 Jan 2005& ATCA& 25.17& \nodata       & \nodata     & 18 $\pm$ 1   & 11 $\pm$ 1 & 5 $\pm$ 1     & 4.3$\pm$ 1 & 2.0$\pm$ 1 & \nodata\\
23.84 Jan 2005& ATCA& 26.94& \nodata       & \nodata     & 12 $\pm$ 1   & 7.9 $\pm$ 1& 5.6 $\pm$ 1   & 3.7$\pm$ 1 & 3.6$\pm$ 1 & \nodata\\
24.85 Jan 2005& ATCA& 27.95& \nodata       & \nodata     & 12 $\pm$ 1   & 11 $\pm$ 1 & 4.2 $\pm$ 1   & 5.1$\pm$ 1 & 3.6$\pm$ 1 & \nodata\\
26.26 Jan 2005& GMRT& 29.36& 104 $\pm$ 31  & 19 $\pm$ 6   &\nodata      & \nodata    & \nodata       & \nodata  & \nodata        & \nodata\\
4.01  Feb 2005& GMRT& 38.14& \nodata       & \nodata      & 10$^{c,d}$  & \nodata    & \nodata       & \nodata  & \nodata        & \nodata\\
16.87 Feb 2005& ATCA& 50.97& \nodata       & \nodata      & 10 $\pm$ 1  & 6.3 $\pm$ 1& 3.3 $\pm$ 0.4 & \nodata  & 2.1 $\pm$ 0.3  & \nodata\\
24.01 Feb 2005& GMRT& 58.11& 28.2 $\pm$ 9  &6.6 $\pm$ 1.4 & \nodata     & \nodata    & \nodata       & \nodata  & \nodata     & \nodata\\
\hline
\end{tabular}
\end{center}
\caption[]{}
\label{tab:fluxes}
\end{table}
\end{landscape}

\begin{table}[ht]
\begin{center}
\setlength{\extrarowheight}{-0.090in}
\begin{tabular}{>{\scriptsize}l >{\scriptsize}c >{\scriptsize}c >{\scriptsize}c >{\scriptsize}c >{\scriptsize}c}
\hline
\hline
Epoch&Beam&Beam PA&Fit Major Axis&Axial Ratio&Fit PA\\
& (mas)&(degrees)& (mas)& & (degrees)\\
\hline
03 Jan 2005 & $349 \times 170$  &  11.8   & 78.2$^{+3.0}_{-2.9}$ & 0.34$^{+0.21}_{-0.34}$ & 54.6$^{+6.7}_{-6.0}$\\
04 Jan 2005 & $593 \times 173$  & $-40.3$  & 72.4$^{+14.5}_{-48}$ & 0.00$^{+0.90}_{-0}$       & 69$^{+20}_{-67}$\\
05 Jan 2005 & $397 \times 168$  & $-25.7$  & 55$^{+18}_{-10}$     & 0.66$^{+0.34}_{-0.66}$   & 74$^{+90}_{-90}$\\
06 Jan 2005 & $329 \times 178$  & $-16.5$  & 75.7$^{+3.0}_{-3.0}$ & 0.48$^{+0.18}_{-0.33}$ & 51.8$^{+9.0}_{-8.8}$\\
07 Jan 2005 & $532 \times 178$  &  40.4  & 78$^{+26}_{-18}$     & 0.60$^{+0.4}_{-0.60}$       & 69$^{+90}_{-90}$\\
10 Jan 2005 & $560 \times 161$  & $-38.8$  & 112$^{+30}_{-42}$    & 0.34$^{+0.33}_{-0.34}$ & 18$^{+28}_{-18}$\\
\hline
\end{tabular}
\end{center}
\caption[]{}
\label{tab:size}
\end{table}

\clearpage
\vspace{0.5cm}
\centerline{\bf SUPPLEMENTAL INFORMATION}
\vspace{0.5cm}

{\bf Observational Details}: {\small 
For the VLA observations we used the standard continuum mode with 2 $\times$ 50\,MHz bands, with
the exception of the 1.46\,GHz observation of 4 January 2005, which was taken in spectral line mode with 8 channels of 
width 3.1\,MHz. We used the extragalactic source 3C\,286 (J1331+305) for flux calibration, 
while the phase was monitored with J1820-254, J1751$-$253, and J1811$-$209. 
The listed flux densities and  uncertainties were measured from the resulting maps by fitting an 
elliptical Gaussian model to the afterglow emission.  
The GMRT observations were performed in dual frequency mode with 16\,MHz bandwidth divided into a
total of 128 frequency channels for 610\,MHz observations, 
and 6\,MHz of bandwidth divided into 64 channels for the 235\,MHz observations.
The observations at 1060 MHz at GMRT were carried out with a
bandwidth of 32 MHz.
3C\,48 (J0137+331) and 3C\,286 were used as flux calibrators, and J1822$-$096 was used as the phase calibrator. 
These sources were also used for bandpass calibration.
We obtained the flux densities of the source by fitting a Gaussian with a background level plus a slope and
removed the contribution from a nearby weak source.
Because of the high density of radio sources in the galactic plane
in which \sgr\ is located, the antenna system noise temperature
has notable contributions from the sky within the telescope beam.
This reduces the signal to noise ratio and an appropriate correction must be
made to the observed flux (especially at low frequencies), since the
flux calibrators which establish the flux scale lie well outside the
galactic plane and are in an environment of less sky temperature.
We applied a $T_{sys}$ correction factor for 3C\,48 of 3.88 and 1.93, and for 3C\,286 of 3.87 and 1.8, for the
235\,MHz and 610\,MHz, respectively.
Both the VLA and GMRT data were reduced and analyzed using the Astronomical Image Processing System.
The ATCA observations were performed in snapshot mode with 100MHz of effective bandwidth.
The amplitude calibrator was J1934$-$638, whereas J1711$-$251, J1817$-$254, and 
J1811$-$209 where used as phase calibrators.
The last of these was observed in a rapid (3 minute) cycle mode to compensate for its poor phase stability.
The flux densities were determined by performing a local parabolic fit to the peak closest to the known position of the source.
The NMA observations were performed at 102\,GHz in D-configuration (the most compact configuration) on 4 January 2005, and in
AB configuration (longest baseline configuration) on 12, 13 January 2005. 
We used NRAO530 for the phase calibration, and assumed it to have a flux density of 2.3\,Jy.\\
}

{\bf Details of Source Size Measurements}: {\small   The source sizes were measured by modeling the calibrated
   visibilities with the model-fitting procedure in DIFMAP.  This
   procedure employs the Levenberg-Marquardt non-linear least
   squares minimization technique while fitting a 6 parameter elliptical Gaussian to
   the visibilities.  
   The errors were determined with DIFWRAP using the following scheme:
   the source size parameters were stepped in small increments around
   their best-fitted value to form a grid of values. At each grid point
   the source size parameters were held fixed while the other model
   parameters were allowed to 'relax' with 4 model-fitting rounds.  
   The 95\% confidence limits were determined by those models that had a
   $\Delta\chi^2 < 12.8$ as measured from the best-fit total $\chi^2$
   (Press, W.~H., Teukolsky, S.~A., Vetterling, W.~T. and Flannery, B.~P.
   {\it Numerical Recipies in C. The art of scientific computing}.
   Cambridge: University Press, 2nd ed. 1992).  
   As a check we used phase only
   self-calibration as well as phase and amplitude self-calibration,
   both of which give consistent source size measurements.  We also
   used 30 second time-averaged data sets (to reduce the number of
   degrees of freedom), and found the best fit model parameters agreed
   to within the errors.\\
}

\begin{table}[ht]
\begin{center}
\setlength{\extrarowheight}{-0.090in}
\begin{tabular}{>{\scriptsize}l >{\scriptsize}c >{\scriptsize}c
>{\scriptsize}c
>{\scriptsize}c >{\scriptsize}c}
\hline
\hline
Frequency&$\alpha_A$&$t_1$&$\alpha_B$&$t_2$&$\alpha_C$\\
(GHz)& &(days)& &(days)& \\
\hline
0.240& $-1.7 \pm 0.1$ & \nodata & \nodata & \nodata & \nodata\\
0.61 & $-1.9 \pm 0.1$ &  \nodata & \nodata & \nodata & \nodata\\
1.4 & $-2.0\pm 0.2$ & $10.7\pm 0.3$ & $-4.1\pm 0.3$ & $13.8\pm 0.2$ & $-0.85 \pm 0.2$\\
2.4 & $-0.95\pm 0.3$ & $9.8\pm 0.2$ & $-3.5\pm 0.2$ & $13.8\pm 0.5$ & $-0.95\pm 0.2$\\
4.9 & $-1.55\pm 0.15$ & $8.8\pm 0.2$ & $-3.1\pm 0.2$ & $18.6\pm 3.0$ & $-0.65\pm 0.3$\\
6.1 & $-2.3 \pm 0.1$ & \nodata & \nodata &\nodata  &\nodata \\
8.5 & $-2.00\pm 0.15$ & $8.1\pm 0.3$ & $-2.8\pm 0.24$ & $18.0\pm 3.0$ & $-0.64\pm0.4$\\
\hline
\end{tabular}
\end{center}
\label{tab:breaks}
\end{table}

{\bf Table~\ref{tab:breaks}}: {\small Summary of temporal indices and breaks in the light curve of \sgr\ in 7 frequency bands.
The fits represent minimums in $\chi^2$ subject to the conditions that the two power-law slopes are continuous at 
the break point and disagree by more than 1-$\sigma$.
The first and second break points are denoted by $t_1$ and $t_2$, respectively.
The temporal decay indices ($S_\nu \propto t^{\alpha_i}$) 
are $\alpha_A$ for $t<t_1$, $\alpha_B$ for $t_1<t<t_2$, and $\alpha_C$ for $t>t_2$.
}

\end{document}